\documentclass[sigconf,natbib=false,balance=true]{acmart}

\usepackage{fancybox}
\usepackage{framed}   %
\usepackage{graphicx}
\usepackage{caption}
\usepackage{subcaption}
\usepackage{hyperref}
\usepackage{listings}

\usepackage[abbreviate=true, dateabbrev=true, isbn=false, doi=true, urldate=comp, url=false, maxbibnames=9, backref=false, backend=biber, style=numeric-comp, language=american, giveninits=true, sortcites=true, sorting=none]{biblatex}

\newcommand\fv[1]{\textcolor{black}{#1}}
\newcommand\mh[1]{\textcolor{black}{#1}}
\newcommand\fr[1]{\textcolor{orange}{#1}}
\renewcommand\fv[1]{#1}
\renewcommand\mh[1]{#1}
\renewcommand\fr[1]{#1}

\definecolor{shadecolor}{gray}{0.9}

\newenvironment{myframed}{%
  \MakeFramed {\advance\hsize-\width \FrameRestore}}%
 {\endMakeFramed}

\acmConference{Accepted for publication in the 29th International Conference on Evaluation and Assessment in Software Engineering (EASE)}{}{17–20 June 2025, Istanbul, Türkiye.}

\copyrightyear{2025}
\acmYear{2025}
\setcopyright{rightsretained}
\acmDOI{}  %
\acmISBN{} %

\makeatletter
\def\@copyrightpermission{
  \thispagestyle{plain}   %
  \hspace*{0mm}\includegraphics[width=2cm]{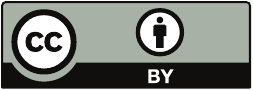}%
  \hspace*{2mm}\raisebox{2.5mm}[25pt][5pt]{%
          \parbox{\columnwidth}{\footnotesize This work is licensed under a Creative Commons \\ Attribution 4.0 International (CC BY 4.0) license.}%
  }%
}%
\makeatother
\begin{document}

\fancyfoot[C]{\footnotesize\thepage} %
\AddToHookNext{shipout/background}{\acmConference[Accepted at EASE 2025]{the 29th International Conference on Evaluation and Assessment in Software Engineering (EASE)}{17–20 June 2025}{Istanbul, Türkiye.}}

\title{%
The Art of Repair: Optimizing Iterative Program Repair with Instruction-Tuned Models%
}

\newcommand{\SimulaAffiliation}{\affiliation{%
  \institution{Simula Research Laboratory}%
  \city{Oslo}%
  \country{Norway}%
}}
\newcommand{\BIAffiliation}{\affiliation{%
  \institution{BI Norwegian Business School}
  \city{Oslo}
  \country{Norway}
}}

\author{Fernando Vallecillos Ruiz}
\email{fernando@simula.no}
\orcid{0000-0001-7213-3732}
\SimulaAffiliation{}
\author{Max Hort}
\orcid{0000-0001-8684-5909}
\email{maxh@simula.no}
\SimulaAffiliation{}
\author{Leon Moonen}
\orcid{0000-0002-1761-6771}
\email{leon.moonen@computer.org}
\SimulaAffiliation{}

\begin{abstract}

\fv{
Automatic program repair (APR) aims at reducing the manual efforts required to identify and fix errors in source code.
Before the rise of Large Language Model (LLM)-based agents, a common strategy was simply to increase the number of generated patches, sometimes to the thousands, which usually yielded better repair results on benchmarks. 
More recently, self-iterative capabilities enabled LLMs to refine patches over multiple rounds guided by feedback.
However, literature often focuses on many iterations and disregards different numbers of outputs.
}

\fv{
We investigate an APR pipeline that  
balances these two approaches, the generation of multiple outputs and multiple rounds of iteration, while imposing a limit of 10 total patches per bug.
We apply three SOTA instruction-tuned LLMs -- DeepSeekCoder-Instruct, Codellama-Instruct, Llama3.1-Instruct -- to the APR task.
We further fine-tune each model on an APR dataset with three sizes (1K, 30K, 65K) and two techniques (Full Fine-Tuning and LoRA), allowing us to assess their repair capabilities on two APR benchmarks: HumanEval-Java and Defects4J.
}

\fv{
Our results show that by using only a fraction (<1\%) of the fine-tuning dataset, we can achieve improvements of up to 78\% in the number of plausible patches generated, challenging prior studies that reported limited gains using Full Fine-Tuning.
However, we find that exceeding certain thresholds leads to diminishing outcomes, likely due to overfitting.
Moreover, we show that base models greatly benefit from creating patches in an iterative fashion rather than generating them all at once. 
In addition, the benefit of iterative strategies 
becomes more pronounced in complex benchmarks.
Even fine-tuned models, while benefiting less from iterations, still gain advantages, particularly on complex benchmarks.
The research underscores the need for balanced APR strategies that combine multi-output generation and iterative refinement. 
}

\end{abstract}


\maketitle

\section{Introduction}

Software bugs are inevitable in the software development cycle, often leading to system failures and increased maintenance costs~\cite{herb2020:cost, odell2017:debugging}.
Automatic Program Repair (APR) aims to reduce the manual effort required to localize and fix errors in source code.
Traditional APR methods can be broadly categorized into pattern-based~\cite{liu2019:avatar, liu2019:tbar, koyuncu2020:fixminer}, heuristic-based~\cite{yuan2020:arja, saha2017:elixir,motwani2023:better}, and constraint-based approaches~\cite{martinez2018:ultralarge, durieux2016:dynamoth}. 
However, they often faced challenges in scalability and adaptability since they tended to not generalize beyond a pre-set group of strategies. 
Consequently, researchers have started to investigate different method to address these limitations.

In recent years, learning-based approaches have tried to address some of these limitations.
Neural Machine Translation (NMT)-based tools~\cite{lutellier2020:coconut, xia2023:automated, xia2022:less} treat program repair as a translation task from buggy code into correct code.
This approach trains on historical bug fixes but it is dependent on the quantity and quality of the data.

Large Language Models (LLMs) have demonstrated promising results in code-related tasks thanks to their extensive pre-training on code repositories.
Models such as CodeLlama~\cite{roziere2023:code} or DeepSeekCoder~\cite{guo2024:deepseekcoder} have shown high competency in code generation, translation, and completion.
The use of LLMs for APR has become an attractive and popular option~\cite{sobania2023:analysis, zhang2024:survey} often outperforming traditional APR methods~\cite{liu2019:avatar, liu2019:tbar, koyuncu2020:fixminer}.
Despite their competency, many LLM-based APR approaches generate hundreds or even thousands of patches for each bug~\cite{xia2023:revisiting, xiang2024:how}.
\fv{
While this approach may improve results,
it increases the computational overhead and can overwhelm developers who must sift through these outputs~\cite{noller2022:trust}.
}

Recently, instruction-tuning~\cite{zhang2024:instruction} has emerged enabling LLMs to follow commands, improving their ability to perform tasks asked by the user. 
Instruction-tuned models have been key in the development of LLM-based agents.
These agents simulate the cycle of debugging by generating patches, executing them, receiving feedback, and refining their previous answers.
They have the potential to improve repair quality by focusing on refining previous answers instead of producing a large quantity of independent patches.
However, many of these works introduce complicated agent-based pipelines that involve intricate control flows and arbitrary components.
Although these systems achieve even higher performance than their non-agentic counterparts, they also introduce overhead and complexity.

\begin{figure}[t]
    \centering
    \includegraphics[width=\linewidth]{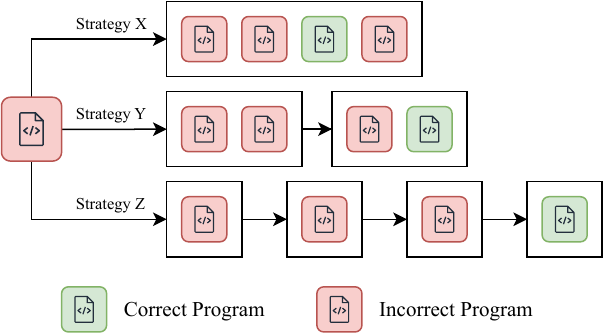}
    \caption{High-level illustration of different APR strategies: (Strategy X) generating a batch of patches all at once; (Strategy Y) iterating over patches in two rounds; (Strategy Z) iterating over one patch over four rounds.}
    \label{fig:overview-strategies}
    \Description[Comparison of three automated program repair workflows displaying batch, two-stage, and iterative single-patch strategies.]{An initial red-bordered program icon branches into three labeled paths called Strategy X, Strategy Y, and Strategy Z. A key beneath the diagram states that green represents a correct program and red an incorrect one. Strategy X leads directly to one rectangle containing several candidate patches, mostly red with one green, indicating a single batch that already includes the correct fix. Strategy Y reaches a first rectangle of red patches, then a second rectangle mixing red and one green, portraying two sequential rounds before success. Strategy Z forms a chain of four rectangles, each holding one patch; the first three are red and the fourth green, illustrating four iterations that test one patch at a time until the fault is removed. }
\end{figure}

\fv{
In the following paper, our goal is to bridge the gap between generating too many independent patches and relying on overly complex iterative pipelines. 
We propose a balanced, \emph{developer-centric} approach limiting the number of generated patches to a practical maximum (e.g., ten patches per bug)~\cite{noller2022:trust}, thereby mirroring real-world constraints while still leveraging iterative refinement. 
}
We assess how the size of fine-tuning data \fv{and technique used} can impact their ability to leverage iterative feedback.
By varying these factors,
we explore whether a larger dataset always leads to better outcomes or if there are caveats.
Additionally, we perform a thorough analysis on different generation strategies (e.g., generating all patches at once, or generating a single patch and iterating over it multiple times) \fv{ as illustrated in Figure~\ref{fig:overview-strategies}}.
By varying the number of iterations and number of outputs per iteration, we aim to maximize repair success while keeping the number of generated patches low, aligning with constraints faced by developers. 

To evaluate our approach, we conduct experiments on two common APR benchmarks: HumanEval-Java~\cite{jiang2023:impact} and Defects4J~\cite{just2014:defects4j}.
For these two datasets, we apply three state-of-the-art instruction-tuned LLMs: Llama 3.1~\cite{dubey2024:llama}, CodeLlama~\cite{roziere2023:code} and DeepSeek-Coder~\cite{guo2024:deepseekcoder}.

The approach proposed focuses on quality over quantity; we are aiming to generate fewer, higher-quality patches so it can be feasible for developers to review them.
The study of agentic pipelines 
provides insights into the optimum generation strategies for iterative tools with the goal of maximizing repair success while keeping the number of generations low.
Our contributions in this work are as follows:
\begin{itemize}
    \item 
    We demonstrate
    the effectiveness of instruction-tuned LLMs for APR 
    \fv{with minimal fine-tuning, challenging prior studies on the limited benefits of FFT and small-scale data.}
    \item We investigate where plausible patches are found within the sequence of generated outputs, providing insights into 
    the contribution of the different positions.
    \item We develop an iterative pipeline enhancing the capabilities of instruction-tuned LLMs to perform APR through consecutive iterations \fv{incorporating execution feedback}.
    \item This work shows how different dataset sizes (1K, 30K, 65K) affect LLM performance on APR, identifying thresholds where further fine-tuning conduces to diminishing results.
    \item We evaluate seven strategies for patch generation, comparing batch generation vs. iterative refinement to determine the optimal trade-off between number of iterations and outputs per iteration.
    \item \mh{We release a replication package with the complete pipeline to ensure reproducibility and facilitate future use.}\footref{foot:figshare-link}
\end{itemize}

The remainder of this paper is organized as follows. Section~\ref{section:background} presents background information on tuning language models as well as related APR approaches.
In Section~\ref{section:design}, we outline our experimental design. 
This includes the research question we pursue as well as studied models and datasets. 
In addition, we describe our implementation details on how we use instruction-tuned models for APR and how to use different iterative strategies to incorporate feedback in the repair process.
The experimental results are shown in Section~\ref{section:results}.
Section~\ref{section:threats} addresses threats to validity and Section~\ref{section:conclusion} concludes our study.

\section{Background and Related Work}
\label{section:background}

\subsection{Fine-tuning LLMs}
Large language models are pre-trained for general purposes, and, although skilled in many tasks, they may not excel in them.
Fine-tuning is a crucial step that allows LLMs to adapt to specialized tasks.
Traditional fine-tuning adjusts all model parameters based on new task-specific data.
As LLMs increase in size, fine-tuning can become computationally costly and susceptible to overfitting~\cite{lu2023:llamareviewer}.
This challenge typically appears in program repair, where collecting diverse high-quality datasets can become difficult.

These issues can be addressed with approaches such as parameter-efficient fine-tuning (PEFT).
These methods propose updating only a subset of the parameters in the fine-tuning process.
One PEFT method for fine-tuning LLMs is Low-Rank Adaptation (LoRA), proposed by Hu et al.~\cite{hu2021:lora}.
LoRA reduces the number of trainable parameters by introducing novel weights for training, rather than updating all weights of the model, as done by Full Fine-tuning (FFT).

In addition to fine-tuning LLMs to specialized tasks, they can be tuned to follow instructions.
Instruction-tuned LLMs have been applied to various software engineering tasks, such as instructional code editing~\cite{cassanoCanItEdit}, program synthesis~\cite{kuznia2022:less,luo2024:semiinstruct} and secure code generation~\cite{he2024:instruction}.
Moreover, models have been trained on different datasets and made publicly available.
Shared instruction-tuned models include the following:
WizardCoder~\cite{luo2023:wizardcoder}, OctoCoder~\cite{muennighoff2024:octopack}, InstructCodeT5+~\cite{wang2023:codet5}, MagiCoder~\cite{weiMagicoderEmpoweringCode}, PanguCoder~\cite{shen2023:pangucoder2}, DeepSeek-Coder~\cite{guo2024:deepseekcoder}, WaveCoder~\cite{yu2024:wavecoder}.
The sharing of models enables an easy reuse and application to various tasks and studies. 

While various works show the capabilities of instruction-tuning on Natural Language Processing (NLP) tasks, Yuan et al.~\cite{yuan2023:evaluating} evaluated 10 open-source models on four code related tasks: defect detection, clone detection, assertion generation, code summarization.
The 10 instruction models have been applied to the different tasks under three settings: zero-shot, few-shot and fine-tuned.
Their findings showed that fine-tuning instruction LLMs can improve performance over zero- and one-shot setting.
Another work, by Zhuo et al.~\cite{zhuo2024:astraios}, studied which PEFT method should be used for instruction-tuning.
In particular, they considered 7 PEFT methods and 4 model sizes of OctoCoder.
Similarly, we set out to investigate the performance of fine-tuning instruction-based models for the program repair task. 
Additionally, we measure the impact of training choices on fine-tuning performance (using LoRA, varying training data sizes). %

\subsection{Iterative refinement for SE}
Successive feedback loops in the generation process have been shown to improve \fv{outcomes}~\cite{chen2023:teaching, liventsev2023:fully}.
This approach is very effective in tasks where initial generations may fail but are close to the desired output.
Each cycle contains additional context, helping to refine the output and adjusting it through successive iterations.

Self-feedback is a common approach in which an LLM generates feedback on its own outputs to refine them without external supervision~\cite{madaan2023:selfrefine}.
Alternatively, a different LLM can be specialized on providing feedback for refinement.
This multi-agentic approach allows for more complex corrections by specializing an LLM on generating concrete helpful feedback to resolve the remaining problems~\cite{jin2024:rgda}.
In addition to feedback from models, external tools like test suites and compilers can also be used in the refinement process.
These tools provide execution feedback, through failed test cases or compilation errors, which the LLM is able to leverage to fix concrete errors in the code~\cite{bouzenia2024:repairagent, hidvegi2024:cigar, ye2024:itera}.
Furthermore, new approaches have been trying to further train LLMs by leveraging execution feedback in real time~\cite{gehring2024:rlef}.

\fv{
Recent work has curated datasets with iterative code-generation goals in mind, featuring multi-turn interaction with human feedback~\cite{zheng2024:opencodeinterpreter}.
However, these datasets rely on artificially introduced errors, GPT-generated bugs, rather than real software faults, which can limit their applicability in true APR scenarios.
In contrast, our research focuses on analyzing how different 
factors
impact iterative repair, yielding new insights into the trade-offs between generation strategies, data size, and fine-tuning techniques.
}

\subsection{LLMs for APR}

Jiang et al.~\cite{jiang2023:impact} investigated four LLMs (PLBART, CodeT5, CodeGen, InCoder) with different number of parameters, and investigated them with and without fine-tuning on a dataset with 143,666 buggy functions and their fix.
Fine-tuning improved the number of repaired programs for every model type and size.
Moreover, they investigated the impact of sharing information of the buggy line in the prompt, which LLMs can learned to make use of after fine-tuning.
They shared their framework for testing LLMs on three APR datasets (Defects4j, HumanEval-Java, Quixbugs). 

Silva et al.~\cite{silva2024:repairllama} proposed the fine-tuned model RepairLLaMA and investigated different combinations of code representations for the input and output for LLMs and their impact on APR performance. 
RepairLLaMA is based on CodeLlama-7b and fine-tuned on the Megadiff dataset~\cite{monperrus2021:megadiff}.
To avoid overfitting, RepairLLaMA was fine-tuned with LoRA, which achieved better performance than full fine-tuning and the use of models without fine-tuning. 
RepairLLaMA was even able to outperform GPT-4 on the Defects4J v2 dataset.

In contrast to conventional APR approaches with LLMs, which generate a range of patches and validate them on tests, one can perform APR in an iterative fashion. 
An example of this are the works by Xia and Zhang~\cite{xia2023:conversational,xia2023:keep} who advocated for conversational APR.
In their work, patches are validated and failing tests are used as feedback to prompt LLMs again with additional information.

The two most closely related works to ours are from Li et al.~\cite{li2024:comprehensive} and Yang et al.~\cite{yang2024:multiobjective}.
Li et al.~\cite{li2024:comprehensive} addressed a lack of APR-specific datasets for instruction tuning. 
Unlike existing datasets to teach language models code instructions, they created the \textsc{APR-Instruction} dataset focused on the single task of APR and the use of enriched instructions. 
Training samples are enriched with problem descriptions and bug causes.
Proceeding, they fine-tuned four pre-trained base LLMs (CodeLLama-7B, CodeLlama-13B, DeepSeek-Coder-6.7B, Llama-2-7B) on this dataset and have evaluated them on three APR datasets (Defects4j, Quixbugs and HumanEval).
Moreover, Li et al. investigated the impact of four PEFT techniques on the performance of fine-tuned models in contrast with full model fine-tuning.
In contrast to this work, which used base LLM models and trained them on an APR instruction-dataset, we further fine-tuned an instruction-model for the APR task.
Thereby, the language models are able to learn one task at a time (first, the base-model is fine-tuned for instructions and then further tuned for APR), rather than multiple tasks at once (instructions and APR).

Yang et al.~\cite{yang2024:multiobjective} proposed a novel multi-objective fine-tuning approach for instruction models on the APR task (\textsc{MORepair}).
\textsc{MORepair} is used to fine-tune instruction LLMs to learn 1) to repair code; 2) provide explanations for the repair.
In addition to the training procedure \textsc{MORepair}, Yang et al. created TUTORLLMCODE, a dataset of 1600 samples with buggy and fixed codes, as well as guidance written by GPT-4.
To fine-tune with fewer parameters, QLoRA is used.
\textsc{MORepair} is applied to four LLM (CodeLlama-13B-instruct, CodeLlama7B-instruct, StarChat-alpha, and Mistral-Instruct7B-v0.1) and tested on new benchmarks (EvalRepair-C++ and EvalRepair-Java), which correspond to HumanEval with additional tests.
Results show that fine-tuning instruct models with \textsc{MORepair} is able to improve performance over base models, standard fine-tuning, as well as existing works which used larger datasets.

\fv{
Although Li et al.~\cite{li2024:comprehensive} and Yang et al.~\cite{yang2024:multiobjective} rely on LLMs to generate their datasets, our approach leverages an existing dataset, rephrasing the samples (buggy and fixed code) to conform with an instruction format.
More importantly, we investigate several repair strategies incorporating execution feedback over multiple rounds, under a practical budget of patches.
This real-world constraint ensures that the pipeline is developer-friendly, in contrast to methods that generate a large number of candidates patches.
In doing so, this work contributes to a deeper exploration on how fine-tuning, iterations, and constraint patch generation interplay to improve APR outcomes.
}

\section{Experimental Design}
\label{section:design}

\subsection{Research Questions}
We address the following research questions in our study:
\begin{enumerate}
    \item[(RQ1)] What impact does fine-tuning instruction models on automatic program repair have on their performance?
    \begin{enumerate}
        \item How does the size of the dataset impact the model's abilities to repair bugs?
        \item Where do models find plausible patches within the sequence of generated outputs?
    \end{enumerate}
\end{enumerate}
To address this question, we fine-tune three instruction-tuned models using subsets of three sizes composed of 1K, 30K, and 65K samples from an APR dataset.
We evaluate the base and fine-tuned models on two APR benchmarks (Section~\ref{section:datasets}).
The evaluation focuses on the LLM ability to generate plausible patches - patches that pass the test suite associated with the problem.
During the evaluation, we do not only analyze the success rate but also the position within the output sequence where plausible patches were found.
We aim to understand the impact of further fine-tuning on the models' repairs by comparing the performance across the different variants of the models and analyze the placement of plausible patches withing the generated outputs.

\begin{enumerate}
    \item[(RQ2)] How does the relationship between the number of outputs per iteration and the total number of iterations influence the effectiveness of APR using LLMs?
    \begin{enumerate}
        \item How do base and fine-tuned models respond to variations in outputs per iteration and total iterations?
        \item What is the optimal combination of outputs per iteration and total iterations with a fixed total output limit?
    \end{enumerate}
\end{enumerate}

To explore this question, we implement multiple generation strategies consisting of different combinations of number of iterations and number of outputs generated per iteration (Section~\ref{section:strategies}).
These strategies are applied to base models and fine-tuned models within our iterative repair pipeline, incorporating feedback in each subsequent iteration.
We again evaluate the success on the same two APR benchmarks as in RQ1.
We assess the success of each strategy by measuring the number of plausible patches generated and further analyze the uniqueness of the problems solved between the strategies.

\subsection{Models}

\textbf{Llama 3}~\cite{dubey2024:llama} is the most recent version of the Llama models~\cite{touvron2023:llama2,touvron2023:llama} created by Meta. 
In contrast to Llama 2, the models are trained on higher quality and quantity of data 
(i.e., 15T tokens as compared to 1.8T tokens for Llama 2).
After pre-training, several rounds of post-training are performed to align Llama 3 with human instructions. This is achieved by supervised fine-tuning and Direct Preference Optimization.
In addition to teaching Llama 3.1 models instructions in a post-processing stage, the models are enhanced with coding capabilities. 
We chose the 8B parameter variant of the Llama 3 model.\footnote{https://huggingface.co/meta-llama/Llama-3.1-8B-Instruct}

\textbf{CodeLlama}~\cite{roziere2023:code} presents a specialized version of Llama 2 for coding tasks. 
In particular, CodeLlama is initialized with Llama 2 models and further trained on 500B tokens from code data present in the Llama 2 training dataset.
Overall, three model types are shared, covering four sizes (7, 13,34, 70B)\footnote{https://huggingface.co/codellama}:
foundation/base models, Python specialized models, instruction-based models.
For instruction tuning, CodeLlama models are fine-tuned on two data sources: 1) proprietary dataset; 2) self-instruct.
The proprietary dataset is based on the instruction-tuning dataset for Llama 2 to teach CodeLlama to follow instructions and abide safety properties. 
The self-instruct dataset consists of interview-style programming questions created by Llama 2 and solutions as well as tests created by CodeLlama.
We chose the 7B parameter variant of the CodeLlama model.\footnote{https://huggingface.co/meta-llama/CodeLlama-7b-Instruct-hf}

\textbf{DeepSeek-Coder}~\cite{guo2024:deepseekcoder} presents a range of open-source code models with 1.3B, 6.7B and 33B parameters. 
The models are based on the DeepSeek LLM architecture~\cite{deepseek-ai2024:deepseek} and are trained from scratch, on 2 trillion tokens from 87 programming languages. 
To allow DeepSeek-Coder to understand instructions, the base models have been fine-tuned on instructions following the Alpaca Instruction format~\cite{taori2023:stanford} and 2B tokens.   
The performance of DeepSeek-Coder models is competitive, being able to perform similar to larger models or even outperforming GPT-3.5 Turbo in several benchmarks.
We chose the 6.7B parameter variant of the DeepSeek-Coder model.\footnote{https://huggingface.co/deepseek-ai/deepseek-coder-6.7b-instruct}

\subsection{Datasets}
\label{section:datasets}
\emph{Fine-tuning:}
The training data for the fine-tuning process is collected from commits of open-source GitHub Java projects~\cite{zhu2021:syntaxguided}.
The dataset contains a total of 143,666 samples of single-hunk fixes.
From this larger dataset, we create three subsets consisting of 1K, 30K, and 65K samples.
These three different subsets allow for the study of the impact of variant scales of fine-tuning data.

\emph{Benchmarks:}
The evaluation of the models is done through two APR benchmarks in Java: Defects4J~\cite{just2014:defects4ja} and HumanEval-Java~\cite{jiang2023:impact}.
Defects4J \fr{v2.0} consists of 835 real-world bugs extracted from open-source Java projects.
We follow the classification of previous work and select a subset of 217 \fr{single-hunk} bugs~\cite{jiang2023:impact}.
On the other hand, HumanEval-Java is a bug benchmark containing 164 single-function bugs.
Due to its recency, it reduces the risk of data leakage in the pre-training.
Both benchmarks contain buggy code and one fix along with unit tests to assess the plausibility of the patches generated.

\subsection{Implementation}
The pipeline for these experiments starts with the problems from the benchmarks.
Each input consists of functions where the buggy code is delimited using the tokens \texttt{<bug\_start>} and \texttt{<bug\_end>}. 
We adopt a beam-based search decoding strategy without stochastic sampling to maintain reproducible and deterministic outputs across all experiments.
The pipeline uses the following template to generate the initial prompt for the LLM:

\begin{verbatim}
"""
The input is buggy code. The bug starts from 
'<bug_start>' and ends at '<bug_end>'. 
Please fix the following code. Return the fixed 
complete method.
```
{buggy_function}
```
"""
\end{verbatim}

The output generated is parsed by looking for the triple backquote symbol (\texttt{\`{}\`{}\`{}}) that may be followed by the keyword (\texttt{\`{}\`{}\`{}java}) to extract the generated code.

In the validation phase, the output generated is inserted into the original problem from the benchmark and the related tests are executed.
This execution results in four possible outcomes:
\begin{itemize}
    \item \textbf{Plausible}: All tests pass.
    \item \textbf{Wrong}: At least one test fails.
    \item \textbf{Timeout}: At least one test times out.
    \item \textbf{Uncompilable}: The generated program cannot be compiled.
\end{itemize}

If the result is plausible, no further processing is done.
The other results extract feedback for the next iteration.
In case of a wrong or timeout result, the pipeline extracts the name of one of the tests that fails or times out.
The pipeline then retrieves the source code of the corresponding test.
If the patch is deemed uncompilable, the pipeline extracts the compilation errors from the logs.
This information is referred to as \textit{feedback} from the validation step.

After the validation step is completed, the pipeline starts the iterative process for the non-plausible patches.
If a patch is marked as \textbf{Wrong} or \textbf{Timeout}, the model is prompted with the previous chat context and extended with the following template:

\begin{verbatim}
"""
The code is still not correct. 
It fails the following test.
```
{failed_test_code}
```
Fix the original code so it passes the test.
"""
\end{verbatim}

If a path is deemed \textbf{Uncompilable}, the following template is used:
\begin{verbatim}
"""
The code is still not correct. It does not compile.
This is the compilation error.
```
{compilation_error}
```
Fix the original code.
"""
\end{verbatim}

This iterative process continues until all problems are solved or the maximum number of patches is generated.

\subsection{Generation Strategies}
\label{section:strategies}
Recent work in automatic program repair generates a prohibiting amount of patches\fv{, sometimes in the thousands,} for every problem~\cite{xia2023:revisiting, xiang2024:how}. 
Generating a large number of patches is resource intensive and further increases the computational cost.
Furthermore, developers were found to be unlikely to consider more than 10 patches~\cite{noller2022:trust}.
When designing the experiments, we take into account the practical limitations of computational resources and evaluation time.
To address these concerns, we are limiting the number of patches generated per bug to a maximum of 10.
This approach not only reduces computational costs, but also aligns with common practices in APR research, allowing comparison with previous work~\cite{silva2024:repairllama,li2024:comprehensive,jiang2023:impact,yang2024:multiobjective}.

We explore different generation strategies by varying the number of outputs in the initial generation ($n_o$), the number of outputs in subsequent generations ($n_i$), and the total number of iterations ($i$).
We ensure that $n_o + (n_i \times i) \leq 10$.

To avoid exponential growth in the number of patches, we focus the iterative process on the first patch generated, as the first patch represents the output LLMs deem most likely to be correct. 
We refine the first patch through iterations rather than iterating on all generated patches. 
This approach ensures that the total number of outputs remains manageable.

\begin{itemize}
    \item \textbf{Strategy A (10$\times$1):} Generate ten outputs in a single iteration.  %
    \item \textbf{Strategy B (8-2):} Generate eight outputs in the first iteration, and two outputs in the next iteration.
    \item \textbf{Strategy C (5$\times$2):} Generate five outputs per iteration over two iterations.
    \item \textbf{Strategy D (6-2-2):} Generate six outputs in the first iteration, and two outputs in the next two iterations.
    \item \textbf{Strategy E (4-3-3):} Generate four outputs in the first iteration, and three outputs in the next two iterations.
    \item \textbf{Strategy F (2$\times$5):} Generate two outputs per iteration over five iterations.
    \item \textbf{Strategy G (1$\times$10):} Generate one output per iteration over ten iterations.

\end{itemize}

\subsection{Evaluation Metric}

\fv{
In our APR pipeline, we evaluate the effectiveness of our approach by measuring the number of problems for which at least one plausible patch is generated.
A patch is plausible if it (1) compiles successfully and (2) passes all tests associated with the targeted bug.
Otherwise, the patch is considered implausible.
We work with 12 models, each generating up to 10 outputs per problem for multiple strategies.
Although some of the outputs may overlap between the strategies, the test suite of each problem allows us to efficiently assess tens of thousands of patches across the experiments.
}

\fv{
\textbf{Manual Assessment and Transparency}:
Across all experiments, we produced over 9,000 plausible patches. 
Given the extensive number of generated patches, manual checking of each one would be prohibitively labor-intensive.
To further confirm correctness, we manually inspected 
\fr{3,298 plausible patches.
Of these, 3,167 were confirmed to be correct, while 131 were found to be overfitting to the test suite.}
To promote transparency and reproducibility, \emph{all} generated patch files, \fr{the code including the seed used to randomly sample these patches,} and manual assessments are released in our replication package, enabling other researchers and practitioners to examine or extend our work.
}

\section{Experimental Results}
\label{section:results}
\subsection{Results of RQ1}
\subsubsection{Influence of data size on fine-tuning}
In RQ1, we investigate the impact of fine-tuning instruction models for APR.
We first focus on the influence of the fine-tuning dataset size.
For that purpose, we present the results in Table \ref{table:hej-10-outputs-comparison} and Table \ref{table:d4j-10-outputs-comparison} with the number of problems with plausible patches by each model variant using Strategy A (i.e., generating 10 outputs in a single iteration).

The results show that fine-tuning, even using a relatively small dataset, enhances the APR performance of LLMs.
For HumanEval-Java, the performance of CodeLlama increased the number of fixed problems to 107 (78\% improvement) after FFT with 1K examples.
Similarly, an improvement of 70\% and 59\% is observed for DeepSeekCoder and Llama3.1 respectively.
Although small data sets led to substantial performance gains, increasing the dataset beyond 1K samples did not always result in improvements.
In some models, performance plateaued or even decreased with larger datasets.
For instance, CodeLlama FFT and DeepSeek-Coder decreased from 107 and 129 problems solved with 1K examples to 100 and 122 problems with 65K examples respectively.
We can observe similar results for the Defects4J datasets. The best performing FFT variants are trained on 1K or 30K samples rather than 65K. 
These findings suggest that there is a threshold beyond which, results diminish.
This problem has been documented for APR~\cite{jiang2023:impact, li2024:comprehensive}.
\begin{myframed}
\noindent
\textbf{Finding 1:} Full fine-tuning can achieve large improvements with relatively small datasets, such as 1K samples.
\end{myframed}
The causes of this can be narrowed down to: 1) Data quality, 2) Overfitting, 3) Limited model capacity.
Li et al.~\cite{li2024:comprehensive} curated a higher quality dataset that suffers from the same problem, therefore lowering the likelihood data quality being the cause.
Moreover, previous work indicates that instruction-tuned models are good Zero-Shot Learners~\cite{wei2022:finetuned} and LLMs are usually undertrained~\cite{hoffmann2022:training} which would suggest that the issue is not due to limited model capacity.
As a result, we believe this problem to be caused by overfitting with large datasets with small degree of variations.
\begin{myframed}
\noindent
\textbf{Finding 2:} Results suggests that there is a threshold beyond which adding more data yields limited 
improvements, 
likely due to overfitting.
\end{myframed}

\begin{table}[t]
\centering
\begin{tabular}{lrrr}
\toprule
 & CodeLlama & DeepSeek-Coder & Llama3.1 \\ \midrule
Base & 60 & 76 & 68 \\ \midrule
FFT (1K) & \textbf{107} & \textbf{129} & 108 \\
FFT (30K) & 104 & 121 & \textbf{113} \\
FFT (65K) & 100 & 122 & 112 \\ \midrule
LoRA (1K) & 75 & 79 & 68 \\
LoRA (30K) & 98 & \textbf{128} & \textbf{118} \\
LoRA (65K) & \textbf{100} & 126 & 109 \\ 
\bottomrule
\end{tabular}
\caption{
\fr{Number of unique problems with at least one plausible patch}
in HumanEval-Java. The best performing training set size is highlighted per model and training regiment (i.e., FFT and LoRA).}
\label{table:hej-10-outputs-comparison}
\end{table}

\begin{table}[t]
\centering
\begin{tabular}{lrrr}
\toprule
 & CodeLlama & DeepSeek-Coder & Llama3.1 \\ \midrule
Base & 31 & 24 & 28 \\ \midrule
FFT (1K) & \textbf{98} & 97 & \textbf{96} \\
FFT (30K) & 85 & \textbf{109} & 85 \\
FFT (65K) & 84 & 104 & 82 \\ \midrule
LoRA (1K) & 32 & 33 & 36 \\
LoRA (30K) & 89 & 81 & 93 \\
LoRA (65K) & \textbf{91} & \textbf{83} & \textbf{103} \\ 
\bottomrule
\end{tabular}
\caption{
\fr{Number of unique problems with at least one plausible patch}
in Defects4J. The best performing training set size is highlighted per model and training regiment (i.e., FFT and LoRA).}
\label{table:d4j-10-outputs-comparison}
\end{table}

Our results using LoRA confirm these findings.
Both, FTT and LoRA methods, achieved substantial performance increases.
However, the number of samples required for each method to achieve successful results differs.
While one thousand samples appear to be sufficient for FFT, the LoRA counterparts do not achieve similar results.
For example, the best results for Defects4J and LoRA are achieved by training on 65K samples for each of the three models.
Previous work on APR focused on PEFT fine-tuning \cite{li2024:comprehensive, silva2024:repairllama} due to promising results compared to FFT.
Li et al.~\cite{li2024:comprehensive} performed an analysis on the size of the fine-tuning dataset only for a PEFT approach after discarding FFT given its lower results after being finetuned on a 30K sample dataset.
Although 
our work 
uses a different dataset, we also achieve similar results on a similar dataset size for a PEFT approach but not on FFT.
Our work shows that similar performance can be achieved with a fraction of the training data if FFT is performed.
\begin{myframed}
\noindent
\textbf{Finding 3:} While prior APR studies saw limited improvements with large datasets using FFT, our results show that it can achieve strong APR performance with less data.
\end{myframed}

\begin{comment}
    \subsection{Result RQ1.1}
Influence of dataset size
\begin{itemize}
    \item Small datasets lead to significant performance gains
    \item Diminishing return with larger datasets

\end{itemize}
Possible reasons
\begin{itemize}
    \item Data quality vs quantity
    \item Overfitting: The other work uses LoRA because FFT decreases performance BUT in our case it happens after a lot more data -> overfitting
    \item Model capacity? Doubt it tho
\end{itemize}
\end{comment}

\subsubsection{Position of plausible patches}

\begin{figure*}[t]
    \centering
    \begin{minipage}{0.96\textwidth}
        \centering
        \includegraphics[width=0.6\linewidth]{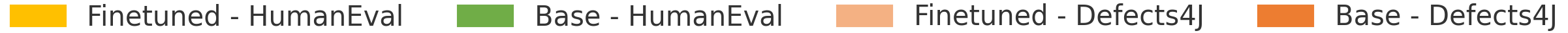}  %
    \end{minipage}

    \begin{subfigure}[b]{0.32\textwidth}
        \centering
        \includegraphics[width=\linewidth]{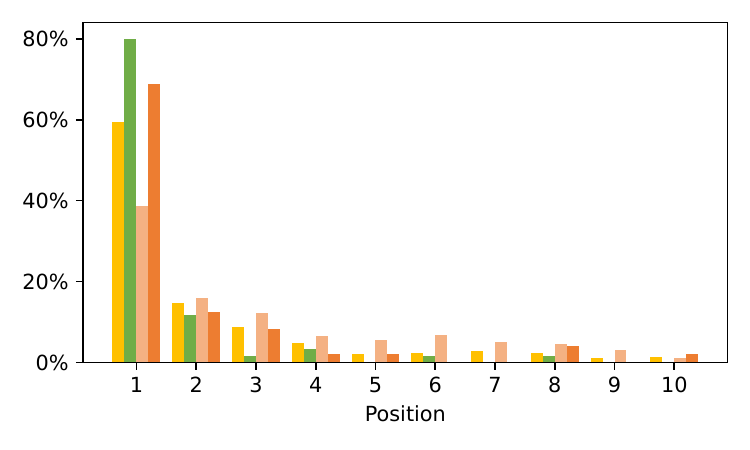}
        \subcaption{CodeLlama}
        \label{fig:a}
    \end{subfigure}
    \hfill
    \begin{subfigure}[b]{0.32\textwidth}
        \centering
        \includegraphics[width=\linewidth]{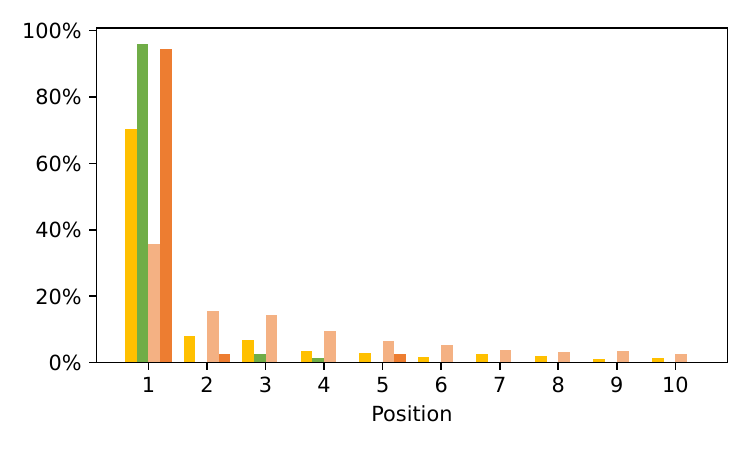}
        \subcaption{DeepSeek-Coder}
        \label{fig:c}
    \end{subfigure}
    \hfill
    \begin{subfigure}[b]{0.32\textwidth}
        \centering
        \includegraphics[width=\linewidth]{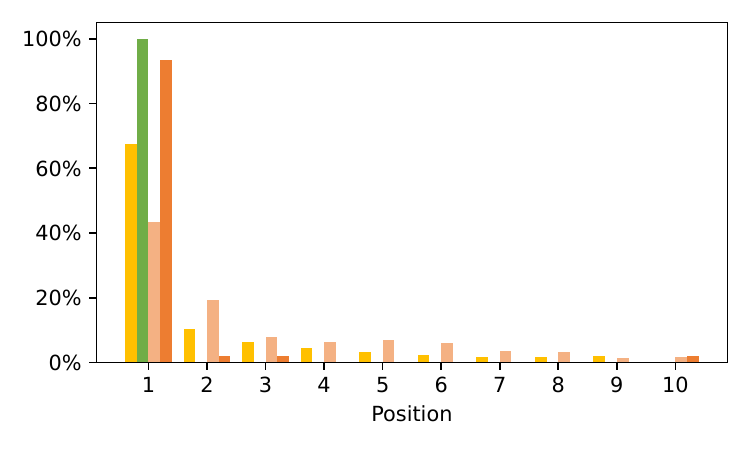}
        \subcaption{Llama3.1}
        \label{fig:e}
    \end{subfigure}
    
    \caption{Position of the first plausible patches found for 10 outputs. Results are shown as proportion for all unique plausible patches found summarized over the 6 fine-tuning configurations for each model. The base model is shown separately.}
    \label{fig:position}
    \Description[Comparison of first plausible patch positions for CodeLlama, DeepSeek-Coder, and Llama3.1 across fine-tuning settings.]{The figure contains three bar plots showing the distribution of the first plausible patch positions for CodeLlama, DeepSeek-Coder, and Llama3.1 models, each labeled as subfigures (a), (b), and (c). For each model, results are shown separately for fine-tuned and base versions evaluated on HumanEval and Defects4J benchmarks, with different colors indicating each condition. All models show that the first plausible patch is most frequently found at position 1, particularly for the base models. Fine-tuned models overall display greater variability, with plausible patches found more often at later positions compared to base models.}
\end{figure*}

We further investigate which of the 10 generated patches, in Strategy A, finds plausible patches. 
Specifically, we count the positions where the first plausible patch for each problem has been found, to show the contribution of each patch position (i.e., 1-10) on the performance of the models. 
Figure~\ref{fig:position} shows these results for the three models on HumanEval-Java and Defects4j. 
For each model, we compare the positions of plausible patches generated by the base models and the fine-tuned models.
For this purpose, we average the results of the six fine-tuned variants (as shown in Table~\ref{table:hej-10-outputs-comparison} and Table~\ref{table:d4j-10-outputs-comparison}).

For HumanEval-Java, we observe that fine-tuned models find the majority of plausible patches in the first five positions. 
In total, we found 90\%, 92\% and 92\% of plausible patches in the first five positions for CodeLlama, DeepSeek-Coder and Llama3.1 respectively. 
When comparing the base model to fine-tuned variants, we can see that the plausible patches generated are focused on the earlier positions, in particular, the first patch accounts for at least 80\% of patches as compared to 60\% for fine-tuned models. 

Plausible patches for Defects4J are more spread out than for HumanEval-Java. 
For instance, the first five patches account for approximately 80\% of the plausible patches found, 10\% less than the first 5 patches for HumanEval-Java. 
Additionally, the performance of the first generated patch by fine-tuned models is almost halved, ranging from 36\% to 43\% for plausible patches found for the three model types.
The base models behave similarly for Defects4J as our observations for HumanEval-Java.
Patches are found earlier than with fine-tuned models, and it is rare to find plausible patches beyond the 5th patch. 

\begin{myframed}
\noindent
\textbf{Finding 4:} 
Base models generate plausible patches in earlier positions with almost no contribution from later patches, while fine-tuned models obtain plausible patches across the first few outputs.
\end{myframed}

Overall, we observe that the later generated patches only contribute a fraction of the whole set of solved problems.
In particular, the last 4-5 out of the 10 generated patches only account for 10\% of the solved problems for fine-tuned models (i.e., 10\% of patches are found in the last four patches for Defects4J and last five patches for HumanEval-Java). 
For base models, there are almost no plausible outputs found in the later patches, with the majority of patches found in the first position.
This leads us to believe that, given a budget of 10 patches for evaluation, one can improve upon the standard practice of generating all 10 patches at once by following an iterative process.  
We investigate this in RQ2.

\begin{myframed}
\noindent
\textbf{Finding 5:} Efforts should be concentrated on early outputs, as later patches contribute less than 10\% to the overall repair success.
\end{myframed}

\subsection{Results of RQ2}
In RQ2, we investigate the influence of fine-tuning on iterative automatic program repair.
In addition, we study the influence of number of iterations and number of outputs per iteration when using iterative LLM-based agents for APR while limiting the number of evaluated patches to 10.
In accordance with RQ1, we start analyzing the results on HumanEval-Java since it prevents data-leakage and consists of simpler cases, and then move onto more complex problems in the Defects4J benchmark.

To limit the models we investigate, we consider a total of three out of the seven studied configurations for each of the LLMs (CodeLlama, DeepSeek-Coder, Llama3.1). 
In particular, we chose the base model as well as the best performing variants trained with FFT and LoRA on HumanEval-Java according to RQ1.

\subsubsection{Influence of fine-tuning on iterations}
\begin{figure*}[t]
    \centering
    \begin{subfigure}{0.48\textwidth}
        \centering
        \includegraphics[width=\textwidth]{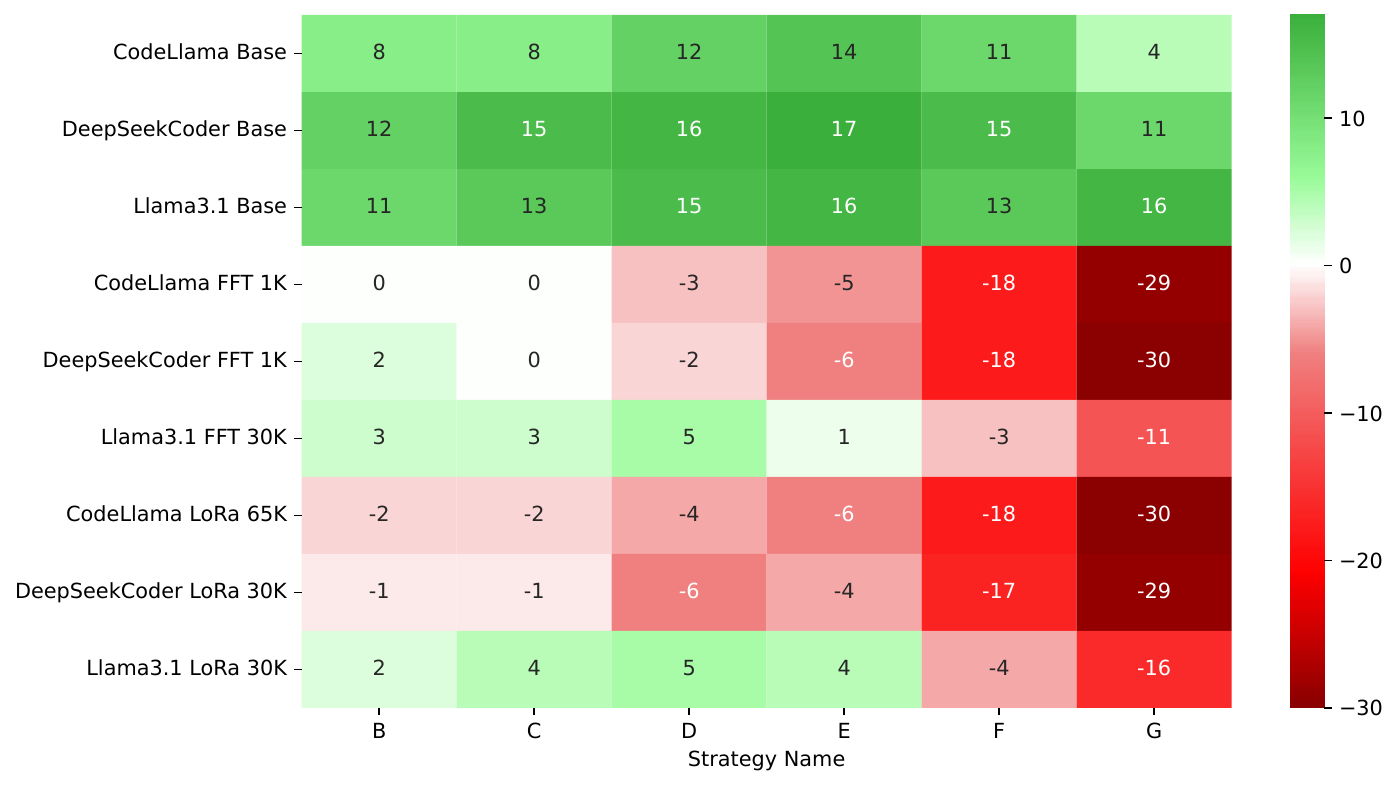}
        \caption{HumanEval-Java}
        \label{fig:humaneval-plausible-compAstrat}
    \end{subfigure}
    \hfill
    \begin{subfigure}{0.48\textwidth}
        \centering
        \includegraphics[width=\textwidth]{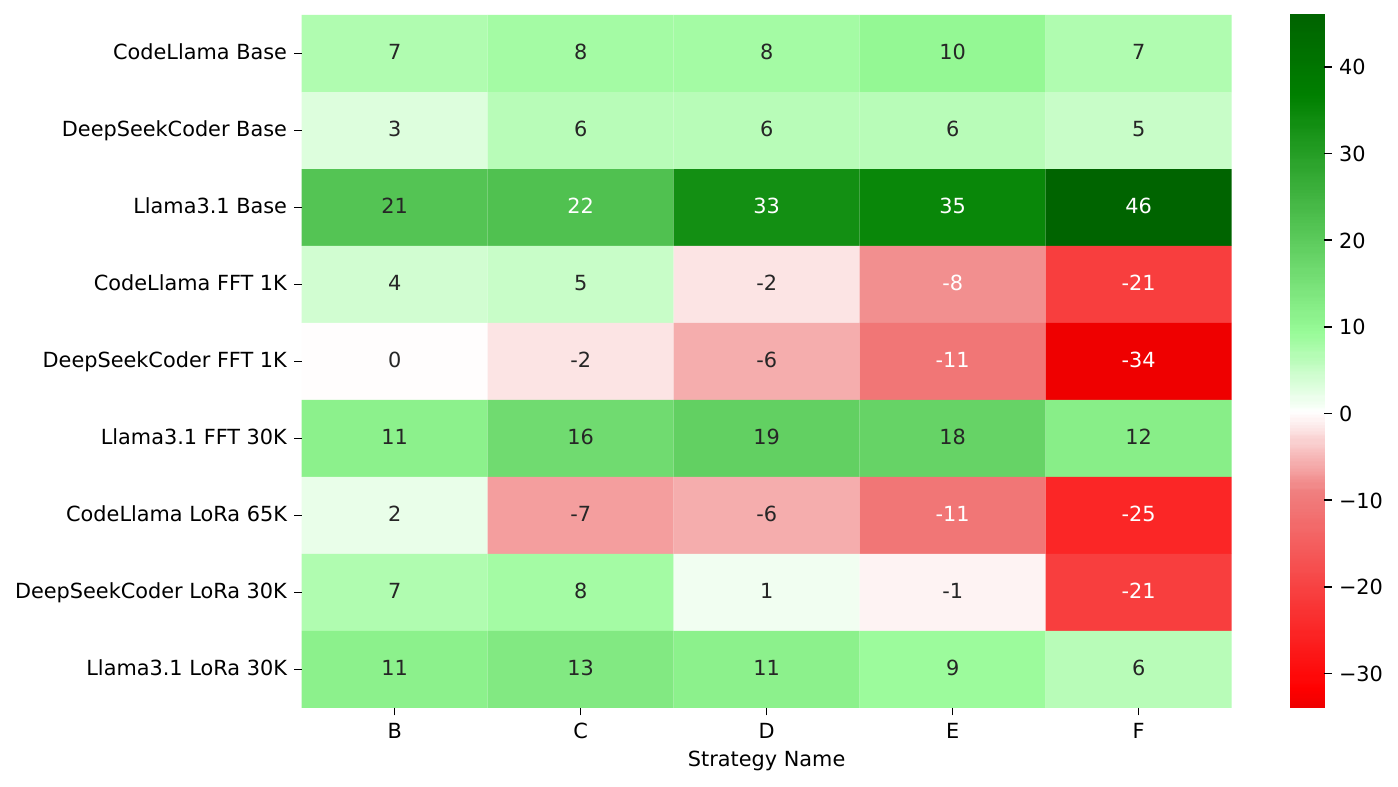}
        \caption{Defects4J}
        \label{fig:d4j-plausible-compAstrat}
    \end{subfigure}
    \caption{Impact of different generation strategies on the 
    \fr{number of unique problems with at least one plausible patch.}
    The heatmaps show the difference in the number of plausible patches with regard to the default strategy (Strategy A).}
    \label{fig:plausible-compAstrat}
    \Description[Impact of generation strategies on the number of plausible patches for HumanEval-Java and Defects4J benchmarks.]{
    The figure contains two heatmaps showing how different generation strategies influence the number of plausible patches compared to a default strategy, labeled A.
    Each row represents a different model variant and each column represents a strategy, labeled B through G. 
    In both heatmaps, green shades indicate an increase in plausible patches and red shades indicate a decrease relative to the default strategy. 
    For HumanEval-Java on the left, base models show consistent increases across strategies, while fine-tuned and LoRa-adapted models often show declines, especially under strategies F and G.
    For Defects4J on the right, Llama3.1 Base shows the largest improvements across all strategies, particularly under strategies E and F, while fine-tuned models such as DeepSeekCoder FFT 1K and CodeLlama LoRa 65K mostly show performance drops. 
    Color intensities correspond to the magnitude of the increase or decrease.}
\end{figure*}

To assess the influence of fine-tuning on the models' ability to integrate iterative feedback, we use two strategies: Strategy A and Strategy G.
We have selected Strategy A as a baseline. %
Strategy A does not iterate and directly generates 10 patches, while strategy G iterates 10 times over the generated patch.
The results are illustrated in Figure \ref{fig:humaneval-plausible-compAstrat} and Figure \ref{fig:d4j-plausible-compAstrat}.

The base models show a consistently improved performance when iterative feedback is incorporated.
This suggests that the feedback is successfully incorporated in successive attempts when failing to generate plausible patches.

In contrast, fine-tuned models achieve their highest performance through fewer iterations.
The number of plausible solutions generated decreased substantially when iterations are added.
This trend is observed across the different models and fine-tuning approaches.
This decreased performance suggests that further fine-tuning of models not only enhances the model's 
initial
solution quality, but it may also reduce their ability to leverage iterative feedback effectively.
A possible explanation is that fine-tuned models quickly become overfitted to a single task.
This would make the models reduce the probability of having diverse outputs in the initial or subsequent iterations.
Alternatively, the fine-tuning process may over-specialize models to perform well under non-iterative conditions, therefore reducing the zero-shot capabilities of the model of incorporating feedback.

\begin{myframed}
\noindent
\textbf{Finding 6:} While iterative feedback consistently improve the performance of base models, fine-tuned models solve a higher number of problems but display reduced effectiveness with iterations, likely due to overfitting.
\end{myframed}

\begin{figure}[t]
    \begin{subfigure}[b]{0.49\textwidth}
        \includegraphics[width=\linewidth]{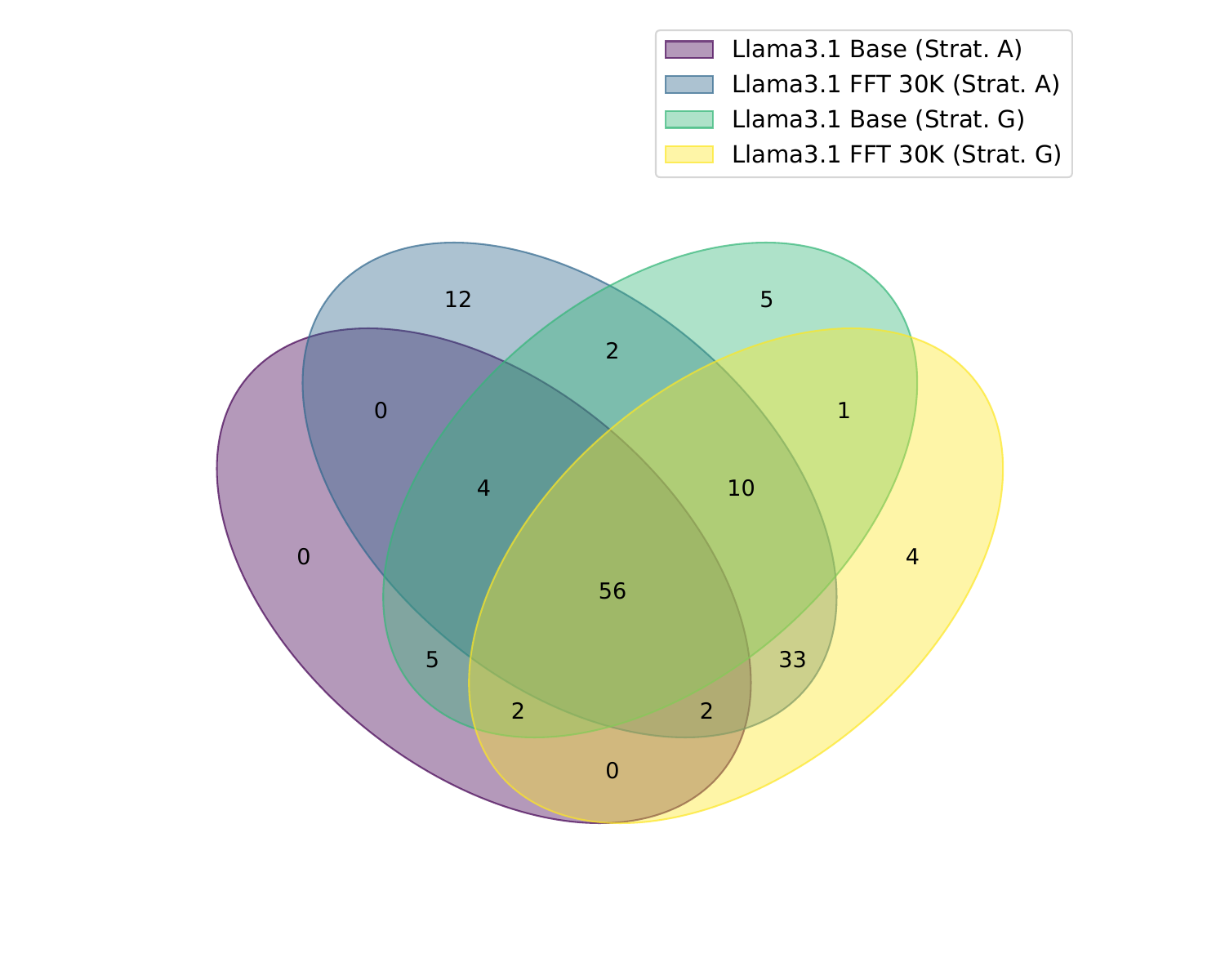}
        \subcaption{HumanEval-Java}\label{fig:hej-venn-diag-llama31}
    \end{subfigure}
    \begin{subfigure}[b]{0.49\textwidth}
        \includegraphics[width=\linewidth]{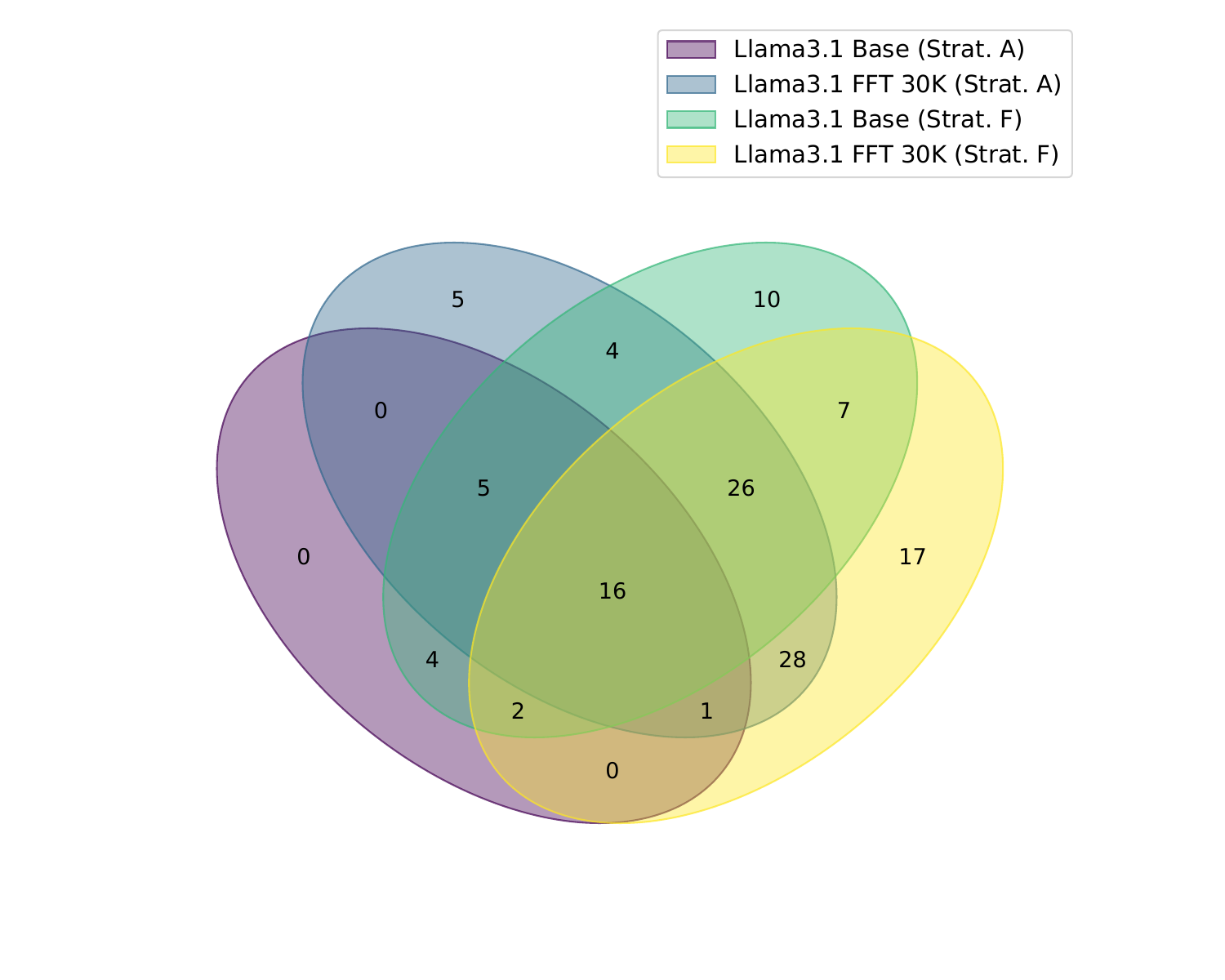}
        \subcaption{Defects4J}\label{fig:d4j-venn-diag-llama31}
    \end{subfigure}
    \caption{Venn diagram of problems with plausible patches generated by the variants of Llama3.1 
    \fr{comparing the least and most iterative strategies applied for each benchmark.}
    }
    \label{fig:general-venn-diag-llama31}
    \Description[Comparison of plausible patch generation by Llama3.1 variants on HumanEval-Java and Defects4J benchmarks.]{Two Venn diagrams display the overlap in problems for which different variants of Llama3.1 generated plausible patches.
    Each diagram shows four model variants with overlapping colored areas. 
    In the HumanEval-Java diagram, a large central overlap of 56 problems is shared by all four models, with smaller regions showing unique or partially shared successes. 
    In the Defects4J diagram, a central overlap of 16 problems is shared across all variants, again surrounded by model-specific and pairwise overlaps. 
    The diagrams illustrate that while there is substantial commonality, each model also solves distinct problems.}
\end{figure}

To further analyze the impact of fine-tuning on the iterative capabilities of the model, we study the problems solved by a base-model and its fine-tuned counterpart.
For simplicity, we have chosen one model, Llama3.1, however, all results are disclosed in the replicability package.
We illustrate the problems solved with the different strategies in Figure \ref{fig:general-venn-diag-llama31}.

The diagram reveals that while the fine-tuned model solves a higher number of problems, there are still problems that only an iterative approach through a base model can solve.
In the case of HumanEval-Java, there are 10 unique problems that Strategy G with a base model can solve that the fine-tuned version are not able to.
In other words, almost $12\%$ of the plausible solutions proposed, are unique to this combination.
This percentage increases to $19\%$ in the Defects4J benchmark when comparing the strategies with the most and least number of iterations.
\begin{myframed}
\noindent
\textbf{Finding 7:} While fine-tuned models solve a greater number of problems, there are certain unique problems that only iterative models can address.
\end{myframed}

This suggest that there are certain problems that inherently benefit from iterative feedback, and the fine-tuned model's ability to solve these is limited.
While fine-tuning greatly enhanced the overall efficiency by improving the initial candidates, it can reduce the ability of the model to leverage iterative feedback.
Therefore, combining fine-tuned models with iterative strategies and designing fine-tuning processes to preserve its zero-shot flexibility would lead to more robust APR models.
To address the limitation of base and fine-tuned models, we investigate how adjustments in the iterative framework can influence APR effectiveness.
\begin{myframed}
\noindent
\textbf{Finding 8:} Combining fine-tuned models with iterative strategies and/or defining fine-tuning processes to preserve zero-shot flexibility may lead to more versatile and robust APR models.
\end{myframed}

\subsubsection{Influence of different strategies}
In addition to the Strategy G, we investigate the remaining iterative strategies, ranging from two to five iterations (see Section~\ref{section:strategies}). 
In terms of the problems solved for each of the strategies, we have identified key trends that are shared between the benchmarks from Figure~\ref{fig:plausible-compAstrat}.

\textbf{Base Models' Performance on HumanEval-Java:}
For base models, increased iterations are beneficial.
Strategies that emphasize fewer outputs per iteration but spread them over multiple ones consistently led to improved results.
However, the highest number of iterations did not always equal better results even for base models.
Strategies like D and E, which balanced the number of iterations and number of outputs per iteration led to the better average performance across the models.
Strategies F and G, which increased iterations even further, led to diminishing returns in some of the models.
This suggests that while iterative refinement often leads to improvement in base models, excessive iterations with minimal output per iteration becomes counterproductive.

\textbf{Fine-Tuned Models' Performance on HumanEval-Java:}
Fine-tuned models, with FFT and LoRA, performed best when more outputs were generated in the initial iterations.
Strategies A and B consistently yield the highest number of plausible patches.
When the number of iterations are increased, the performance quickly decreased.
The decrease indicates that fine-tuned models have a high probability of producing plausible patches early on.
Additionally, their ability to incorporate feedback is substantially reduced compared to base models, in accordance with previous sections.

\begin{myframed}
\noindent
\textbf{Finding 9:} Base models benefit from iterative strategies, but reach a point of diminishing returns when iterations are increased too much.
\end{myframed}

\textbf{Resource Allocation Implications:}
The selected strategy has strong implication on the computational efficiency of the generation process.
Generating 10 outputs in the same inference is less computationally intensive than obtaining 10 outputs via 10 consecutive iterations.
Furthermore, when incorporating feedback in each new iteration, the GPU memory needed for the inference increases.
Given the length of the problems in the Defects4J benchmark, the inference for one single problem may take up to 200GB of GPU memory after repeated iterations.
While GPU memory may be a constraint in certain applications, testing can become a bottleneck in some contexts.
Therefore, the selection of the strategies is always depending on the context and the specific bottlenecks present.
Due to these considerations, we decided that the resource requirement for increased iterations outweighed the marginal gains in performance.
Consequently, we decide to drop Strategy G for the Defects4J benchmark.
This decision allowed us to still provide a balanced view of the different strategies without excessive GPU memory demands.

\begin{myframed}
\noindent
\textbf{Finding 10:} Increasing the number of outputs per iteration while minimizing the number of total iterations reduces the peak memory usage, making certain generation strategies more efficient if testing is not a constraint.
\end{myframed}

\textbf{Base Models' Performance on Defects4J:} 
We see similar trends where more iterations lead to improved performance for base models.
For instance, Llama3.1 Base increases its number of plausible patches from 28 under Strategy A to 74 under Strategy F.
This trend suggests that the complexity of Defects4J may be amplifying the benefits of iterative refinement.
Similar to the previous benchmark, the other base models do not achieve their highest number plausible patches with the highest number of iterations, reemphasizing that each model may have different optimum strategies.

\textbf{Fine-Tuned Models' Performance on Defects4J:} 
\fv{
A similar trend appears for fine-tuned models, where an increase in iterations does not always result in improved outcomes.
In some cases, performance declines as iterations increase, with the negative impact becoming more and more pronounced.
}
For example, DeepSeekCoder FFT 1K generates 97 plausible patches under Strategy A, but drops to 63 under Strategy ~F.
CodeLlama and DeepSeekCoder still present improved performance with low number of iterations.
However, Llama3.1 shows an improvement with a higher number of iterations, but its peaked performance is reached in Strategy C and D.
These results show that even fine-tuned models may still benefit from iterations if the problems are complex enough to require them.

\textbf{Contrasting Trends Between Benchmarks:}
We have showcased that different strategies are optimal for each benchmark.
On the simpler and straightforward HumanEval-Java benchmark, base models showed improvement while fine-tuned models do not benefit from increased iterations.
On the other hand, Defects4J consists of challenging and complex problems that benefit from iterative strategies.
Base models significantly improved with more iterations, while fine-tuned models peaked earlier.
These results emphasize that, while most fine-tuned models can address simpler bugs with minimal iterations, base models can obtain more substantial gains from the iterative refinement.

\begin{myframed}
\noindent
\textbf{Finding 11:} Complex problems benefit greatly from iterative refinement, but each model has an optimal iteration threshold after which gains decline.
\end{myframed}

\section{Threats to validity}
\label{section:threats}
One key internal threat is the potential data leakage from the benchmarks into the pretraining data of the models\fr{, specially for older widely-known benchmarks like Defects4J}.
We mitigate this threat by assessing the models on a \fr{complementary} recent benchmark specifically designed to address data leakage (i.e., HumanEval-Java)\fr{, and by including models like Llama3.1 which have shown less susceptibility to memorization \cite{ramos2025:are}.}
Another internal validity threat is the introduction of bias when selecting hyperparameters in the fine-tuning and generation.
We address this by using standardized hyperparameters and generation setting across the different models\fv{, thus reducing performance variability attributable to tuning decisions.
}

The main external threat is the use of two benchmarks on the same programming language, which may limit the applicability of our approach.
To mitigate this threat, we select benchmarks that are widely known in the literature and cover a range of real-world bug types.
The insights in this paper should generalize to arbitrary programming languages.

\fv{
A key construct validity threat arises from our reliance on plausibility, a binary metric where a patch is considered successful if it compiles and passes all the tests.
Although practical, this metric does not guarantee true correctness of the patch, since tests may not cover all edge cases and outputs may overfit to the tests.
To partially address this limitation, we have manually analyzed over 3,000 of the 9,000 plausible patches generated across our experiments.
Manual checks are labor-intensive and subjective, since reviewers may apply different standards~\cite{wang2020:automated}.
We have published our manual assessments alongside the patches in our replication package.\footref{foot:figshare-link}
This step promotes transparency and has the potential to reduce future validation efforts.
Another construct threat relates to the stopping criteria.
}
The maximum number of iterations or outputs per iteration could impact the evaluation.
To mitigate this threat we selected the fixed number of outputs based on empirical observations from related work~\cite{silva2024:repairllama,li2024:comprehensive,jiang2023:impact,yang2024:multiobjective}.

\section{Conclusion}
\label{section:conclusion}
In this work, we investigated the effectiveness of instruction-tuned LLMs in APR tasks through an iterative repair pipeline.
We focused on three state-of-the-art models, CodeLlama, DeepSeekCoder, and Llama3.1.
With the help of these models, we studied the position of the plausible patches within the generated batch.
We explored how fine-tuning said models with varying size of APR datasets impacts their effectiveness.
This work study the different generation strategies to balance the number of iterations and the number of outputs per iteration.

Our experiments on two widely recognized APR benchmark, HumanEval-Java and Defects4J, revealed that FFT with relatively small datasets can lead to substantial improvements in repair performance.
This suggest that limiting the amount of data in the fine-tuning can enhance the model's ability to generate plausible patches.
In other words, increasing the fine-tuning dataset size did not consistently yield better results and, in some cases, led to declined performance.
Moreover, the analysis of different generation strategies showed that base models benefit from an increased number of iterations.
Yet, we find also an upper threshold for the number of iterations beyond which the process became counterproductive due to excessive resource demands with diminishing returns.
On the other hand, fine-tuned models generated plausible patches in earlier iterations and benefited from less iterative feedback, especially on simpler tasks.
The optimal generation strategy varies depending on the bottlenecks of the process and the complexity of the task:
while fine-tuned models achieved their best results with fewer iterations, base models improved their outcomes substantially with higher number of iterations. %

Resource efficiency was a crucial factor in our study.
By focusing on generating fewer patches, our approach aligns with real-world developer constraints and enhances computational efficiency.
While strategies that minimize iterations reduced memory usage, incorporating feedback in subsequent iterations was able to generate plausible patches for unique problems that fine-tuned models could not.
Our emphasis on resource-efficient generation ensures that the benefits of APR using LLMs can be put into practice without imposing excessive computational costs.

This study contributes with insights into the optimization of instruction-tuned LLMs for APR tasks.
The work presented highlights the importance of selecting an appropriate generation strategy based on the fine-tuning specification of the models and the complexity of the problem.
By displaying that significant performance gains can be achieved with small datasets and suitable generation strategies, we present a path towards accessible and efficient APR.
This work does not only advance the field of APR, but it also provides practical guidance for the deployment of LLMs in complex pipelines.

\subsection{Future Work}
This study opens multiple avenues for future research.
Methods and regularization to mitigate overfitting when fine-tuning with larger datasets could further improve model performance.
Additionally, exploring hybrid generation strategies that can dynamically adapt based on the problem complexity and feedback incorporated can lead to enhanced repair effectiveness.

Incorporating different types of feedback, such as from developers, and integrating these models into development tools can narrow the gap between research and industry.
By continuing to refine models and methodologies in an efficient manner, we can move closer to realizing the potential of automatic program repair, contributing to the end-goal of reliable and efficient software development process.

\section{Data Availability}
The replicability package for this work is available online.\footnote{\url{https://doi.org/10.5281/zenodo.15294695}\label{foot:figshare-link}}
This package includes (1) the source code required to replicate the experiments presented in this work, (2) the generated patches for both benchmarks by all models used, and (3) the nine fine-tuned models.

\section*{Acknowledgments}

This work is supported by the Research Council of Norway through the secureIT project (IKTPLUSS \#288787), and by the European Union through the Horizon Europe Marie Sk\l{}odowska-Curie Actions (\#101151798).
The empirical evaluation made use of the Experimental Infrastructure for Exploration of Exascale Computing (eX3), 
financially supported by the Research Council of Norway under contract \#270053. 
In addition, we acknowledge Sigma2, Norway for awarding this project access to the LUMI supercomputer, owned by the EuroHPC Joint Undertaking, hosted by CSC (Finland) and the LUMI consortium through the Research Council of Norway. 

\printbibliography

 \end{document}